\documentstyle[sprocl,bbox,epsf]{article}
\def\NPA{{\em Nucl. Phys.} A }

\def\PLB{{\em Phys. Lett.}  B }
\def\PRL{{\em Phys. Rev. Lett.} }
\def\PR{{\em Phys. Rev.} }
\def\PRC{{\em Phys. Rev.} C }
\def\PRD{{\em Phys. Rev.} D }
\def\ZPC{{\em Z. Phys.} C }
\def\EPJC{{\em Eur. Phys. J.} C }
\def\half{{\textstyle{1\over 2}}}
\def\be{\begin{equation}}
\def\ee{\end{equation}}
\def\bea{\begin{eqnarray}}
\def\eea{\end{eqnarray}}

\begin{document}


\title{HANBURY BROWN -- TWISS INTERFEROMETRY \\
       IN HIGH ENERGY NUCLEAR AND PARTICLE PHYSICS}

\author{Ulrich Heinz}

\address{Institut f\"ur Theoretische Physik, Universit\"at
  Regensburg,\\
  D-93040 Regensburg, Germany}


\maketitle\abstracts{
 I review recent applications of two-particle intensity interferometry
 in high energy physics, concentrating on relativistic heavy ion
 collisions. By measuring hadronic single-particle spectra and
 two-particle correlations in hadron-hadron or heavy-ion collisions,
 the size and dynamical state of the collision fireball at freeze-out
 can be reconstructed. I discuss the relevant theoretical methods and
 their limitations. By applying the formalism to recent pion
 correlation data from Pb+Pb collisions at CERN we demonstrate that
 the collision zone has undergone strong transverse growth before
 freeze-out (by a factor 2 in each direction), and that it expands
 both longitudinally and transversally. From the thermal and flow
 energy density at freeze-out the energy density at the onset of
 transverse expansion can be estimated from conservation laws. It
 comfortably exceeds the critical value for the transition to color
 deconfined matter.  
}

\section{Introduction}
\label{sec1}

The principle of two-particle intensity interferometry, developed by
R. Hanbury Brown and R.Q. Twiss in the mid 1950's for radio and
optical astronomy,\cite{HBT} was independently rediscovered by
particle physicists just a few years later.\cite{GGLP} It has since
become known under the names ``HBT interferometry'', ``GGLP
effect'', and ``Bose-Einstein correlations''. The method exploits the
effects on the phase-space density of Bose-Einstein symmetrization (or
Pauli antisymmetrization) of multiparticle states of identical
particles. In astronomy one studies photon pairs, in nuclear and 
particle physics one typically uses pion, kaon or nucleon pairs
(recently also pion triplets.\cite{3pi,HZ97}) Strictly speaking,
the method is used oppositely in astronomy and high energy physics:
in astronomy one measures the two-photon correlation
function as a function of the {\em space-time} distance of the two
detected photons and extracts from it information about the size of
the emitter in momentum space (specifically in the opening angle
between the two photon momentum vectors). By knowing the distance of
the emitter (star, radio source) this ``angular size'' in momentum
space can then be translated into a spatial radius of the source. In
particle physics one measures the correlation function as a function
of the {\em momentum} difference between the two pions and extracts
from it information about the space-time extension of the emitting
source. These two opposite ways of looking at intensity interferometry
emphasize its generic nature as a phase-space effect which can only be
understood if both the space-time and momentum-space structure of the
emitter are taken into account simultaneously.
 
Another big difference between HBT interferometry in astronomy and
nuclear physics is that the sources studied in astronomy are
static on the timescale of the observation, while the sources 
created in high energy collisions are highly dynamical and very
shortlived. As I will show this has dramatic implications for the HBT
formalism to be used for the analysis of measured correlation
functions which were only gradually realized during the last 10 years
and for which a common understanding was developed only quite recently.

Relativistic heavy ion collisions are performed in order to create 
color deconfined strongly interacting matter, the ``quark-gluon
plasma''. This plasma is hot and generates a huge pressure which
drives a strong collective expansion of the reaction zone into the
surrounding vacuum. As a consequence of expansion, the quark-gluon
matter cools and dilutes until it can no longer remain in the color
deconfined state: it hadronizes. If the reaction zone is sufficiently
large and locally equilibrated, this process manifests itself as a
bulk phase transition, very similar to the quark-hadron phase
transition in the very early universe (about 20-40 $\mu$s after the
Big Bang). Two crucial questions are therefore (a) whether the
initial energy density in the reaction region is large enough to make
color deconfined quark-gluon matter, and (b) whether this partonic
system equilibrates sufficiently to undergo a thermodynamic phase
transition while expanding. While the longitudinal expansion of the
reaction zone  may be simply due to incomplete stopping of the two 
colliding nuclei, {\em transverse} collective expansion flow can only
be generated by the buildup of a locally isotropic pressure component.
Since this requires a certain degree of local equilibration of the
momentum distributions, the identification of collective {\em
  transverse} flow, superimposed by random thermal motion, plays an
essential role in any attempt to answer these two questions. As we
will see, two-particle interferometry features prominently in such an
endeavour.   
 
Over the last years ample evidence was accumulated that the hot and
dense collision region in relativistic heavy ion collisions indeed 
thermalizes and shows collective dynamical behaviour. Most of this 
evidence is based on a comprehensive analysis of the hadronic single 
particle spectra. It was shown that all available data on hadron 
production in heavy ion collisions at the AGS and the SPS can be 
understood within a simple model which assumes locally thermalized 
momentum distributions at freeze-out, superimposed by collective 
hydrodynamical expansion in both the longitudinal and transverse 
directions.\cite{LH89,Stach94} The collective dynamical behaviour in 
the transverse direction is reflected by a characteristic dependence 
of the inverse slope parameters of the $m_\perp$-spectra (``effective 
temperatures'') at small $m_\perp$ on the hadron masses.\cite{LH89} 
New data from the Au+Au and Pb+Pb systems~\cite{QM96} support 
this picture and show that the transverse collective dynamics is much 
more strongly exhibited in larger collision systems than in the 
smaller ones from the first rounds of experiments. The amount of 
transverse flow also appears to increase monotonically with collision 
energy from GSI/SIS to AGS energies, but may show signs of saturation 
at the even higher SPS energy.\cite{QM96}  

The extraction of flow velocities and thermal freeze-out temperatures 
from the measured single particle spectra relies heavily on model 
assumptions.\cite{LH89} The single-particle spectra are ambiguous 
because they contain no direct information on the space-time structure 
and the space-momentum correlations induced by collective flow. In 
terms of the phase-space density at freeze-out (``emission function'') 
$S(x,p)$ the single-particle spectrum is given by $E\, dN/d^3p = \int 
d^4x\, S(x,p)$; the space-time information in $S$ is completely washed 
out by integration. Thus, on the single-particle level, comprehensive 
model studies are required to show that a simple hydrodynamical model 
with only a few thermodynamic and collective parameters can fit all 
the data, and additional consistency checks are needed to show that 
the extracted fit parameter values lead to an internally consistent 
theoretical picture. The published literature abounds with examples 
demonstrating that without such consistency checks the theoretical 
ambiguity of the single particle spectra is nearly infinite.  

At this point Bose-Einstein correlations between the momenta 
of identical particle pairs provide crucial new input. They give 
direct access to the space-time structure of the source {\it and}
its collective dynamics. In spite of some remaining model 
dependence, the set of possible model sources can thus be reduced 
dramatically. The two-particle correlation function $C(\bbox{q,K})$ is 
usually well approximated by a Gaussian in the relative momentum $q$ 
whose width parameters are called ``HBT (Hanbury~Brown-Twiss) radii''.  
It was recently shown~\cite{HB95,CSH95,CNH95} that these radius 
parameters measure certain combinations of the second central 
space-time moments of the source. In general they mix the spatial and 
temporal structure of the source in a nontrivial way,\cite{CSH95} and 
the remaining model dependence enters when trying to unfold these 
aspects.  

Collective dynamics of the source leads to a dependence of the HBT 
radii on the pair momentum $K$.\cite{P84,MS88} This feature was
recently quantitatively reanalyzed, both
analytically~\cite{CSH95,CNH95,AS95,CL96} and
numerically.\cite{WSH96,HTWW96} The velocity gradients associated 
with collective expansion lead to a dynamical decoupling of different 
source regions, and the HBT radii measure the size of the resulting
``space-time regions of homogeneity'' of the source~\cite{MS88,AS95}
around the point of maximum emissivity for particles with the measured
momentum $K$. The velocity gradients are smeared out by random thermal
motion of the emitters around the fluid velocity.\cite{CSH95} Due to 
the exponential decrease of the Maxwell distribution, this thermal
smearing factor shrinks with increasing transverse pair momentum
$K_\perp$; this is the basic reason for the $K_\perp$-dependence of 
the HBT radii. 

Unfortunately, other gradients in the source (for example spatial and 
temporal temperature gradients) can also generate a $K$-dependence of 
the HBT radii.\cite{CSH95,CL96,TH98a} Furthermore, the pion spectra in 
particular are affected by resonance decay contributions, but only at 
small $K_\perp$. This may also affect the HBT radii in a 
$K_\perp$-dependent way.\cite{Schlei,WH96} The isolation of 
collective flow, in particular transverse flow, from the 
$K_\perp$-dependence of the HBT radii thus requires a careful
investigation of these different effects.  

Our group studied this $K$-dependence of the HBT radii within a simple 
analytical model for a finite thermalized source which expands both 
longitudinally and transversally. For presentation I use the 
Yano-Koonin-Podgoretskii (YKP) parametrization of the correlator 
which, for sources with dominant longitudinal expansion, provides an 
optimal separation of the spatial and temporal aspects of the
source.\cite{CNH95,HTWW96} The YKP radius parameters are independent
of the longitudinal velocity of the observer frame. Furthermore, in all 
thermal models without transverse collective flow, they show perfect 
$M_\perp$-scaling (in the absence of resonance decay contributions). 
Only the transverse gradients induced by a non-zero transverse flow 
can break this $M_\perp$-scaling, causing an explicit dependence on 
the particle rest mass. This allows for a rather model-independent 
identification of transverse flow from accurate measurements of the 
YKP correlation radii for pions and kaons. High-quality data should 
also allow to control the effects from resonance decays.  

A comprehensive and didactical discussion of the formalism and a more
extensive selection of numerical examples can be found in the lecture
notes~\cite{He96} to which I refer the interested reader for more
details. 

\section{Spectra and emission function}
\label{sec2}
\subsection{Single-particle spectra and two-particle correlations}
\label{sec2.1}

The covariant 1- and 2-particle momentum spectra are defined by
 \begin{eqnarray}
   P_1(\bbox{p}) 
  & = & E\, \frac{dN}{d^3p} 
        = E \, \langle\hat{a}^+_{\bbox{p}} \hat{a}_{\bbox{p}}\rangle \, ,
 \label{1} \\
   P_2(\bbox{p}_a,\bbox{p}_b) 
  & = & E_a\, E_b\, \frac{dN}{d^3p_a d^3p_b}
        = E_a \, E_b\, 
          \langle\hat{a}^+_{\bbox{p}_a} \hat{a}^+_{\bbox{p}_b}
                 \hat{a}_{\bbox{p}_b} \hat{a}_{\bbox{p}_a} \rangle \, ,
 \label{2}
 \end{eqnarray}
where $\hat{a}^+_{\bbox{p}}$ ($\hat{a}_{\bbox{p}}$) creates (destroys) a
particle with momentum $\bbox{p}$. They are normalized to $\langle N
\rangle$ and $\langle N(N-1)\rangle$ (i.e.the average number of pions
or of pions in pairs per event), respectively. The angular brackets
denote an ensemble average $\langle \hat O \rangle = {\rm tr}\, (\hat
\rho \hat O)$ where $\hat \rho$ is the density operator associated
with the ensemble. The two-particle correlation function is defined as  
 \begin{equation}
 \label{3}
   C(\bbox{p}_a,\bbox{p}_b)
   = \frac{P_2(\bbox{p}_a,\bbox{p}_b)}{P_1(\bbox{p}_a)P_1(\bbox{p}_b)} \, .
 \end{equation}
If the two particles are emitted independently and final state 
interactions are neglected one can prove~\cite{He96} a generalized
Wick theorem
 \begin{equation}
 \label{corr}
  C(\bbox{p}_a, \bbox{p}_b) = 1 \pm 
  {\vert \langle \hat a^+_{\bbox{p}_a} \hat a_{\bbox{p}_b} \rangle \vert^2
   \over
   \langle \hat a^+_{\bbox{p}_a} \hat a_{\bbox{p}_a} \rangle 
   \langle \hat a^+_{\bbox{p}_b} \hat a_{\bbox{p}_b} \rangle } \, .
 \end{equation}
Note that the second term is positive definite, i.e. the correlation 
function cannot, for example, oscillate around unity. This is no
longer true if final state interactions are included (see below).
I will here assume that the emitted particles are bosons which I will
call pions.

\subsection{Source Wigner function and spectra} 
\label{sec2.2}

In the language of the covariant current formalism~\cite{GKW79}
the source of the emitted pions can be described in terms of classical
currents $J(x)$ which act as classical sources of freely propagating
pions. They parametrize the last collision from which the free
outgoing pion emerges. Very helpful for the following will be the
so-called ``emission function" $S(x,K)$:~\cite{P84,S73} 
 \begin{equation}
   S(x,K) = \int\frac{d^4y}{2(2\pi)^3}\, e^{-iK{\cdot}y}
   \left\langle J^*(x+\half y)J(x-\half y)\right\rangle \, .
 \label{8f}
 \end{equation}
It is the Wigner transform of the density matrix associated with the 
classical source amplitudes $J(x)$. This Wigner density is a quantum 
mechanical object defined in phase-space $(x,K)$; it is real but not
always positive definite. Textbooks on Wigner functions show that
their non-positivity is a genuine quantum effect resulting from the
uncertainty relation and is concentrated at short phase-space
distances; when the Wigner function is averaged over phase-space
volumes large compared to the volume $(2\pi\hbar)^3$ of an
elementary phase-space cell, the result is real and positive definite
and behaves exactly like a classical phase-space density.  

The emission function $S(x,K)$ is thus the quantum mechanical analogue
of the classical phase-space distribution which gives the probability
of finding at point $x$ a source which emits free pions with momentum 
$K$. It allows to express the single-particle spectra and two-particle
correlation function via the following fundamental
relations:~\cite{P84,S73,CH94} 
 \begin{eqnarray}
  E_p {dN \over d^3p} &=&
  \int d^4x\, S(x,p) \, ,
 \label{spectrum}\\
  C(\bbox{q},\bbox{K}) &=& 1 + 
  {\left\vert \int d^4x\, S(x,K)\, e^{iq{\cdot}x} \right\vert^2
   \over
   \int d^4x\, S(x,K+\half q) \ \int d^4x\, S(x,K-\half q)}\, .
 \label{correlator}
 \end{eqnarray}
For the single-particle spectrum (\ref{spectrum}), the Wigner function
$S(x,p)$ on the r.h.s. must be evaluated on-shell, i.e. at $p^0=E_p
= \sqrt{m^2 + \bbox{p}^2}$. For the correlator (\ref{correlator}) we 
have defined the relative momentum $\bbox{q} = \bbox{p}_a - 
\bbox{p}_b$, $q^0 = E_a-E_b$ between the two particles in the pair, 
and the total momentum of the pair $\bbox{K} = (\bbox{p}_a + 
\bbox{p}_b)/2$, $K^0= (E_a+E_b)/2$. Of course, since the 4-momenta
$p_{a,b}$ of the two measured particles are on-shell, $p^0_i = E_i =
\sqrt{m^2 + \bbox{p}_i^2}$, the 4-momenta $q$ and $K$ are in general 
off-shell. They satisfy the orthogonality relation
 \begin{equation}
 \label{ortho}
   q \cdot K = m_a^2-m_b^2=0\,.
 \end{equation}
Thus, the Wigner function on the r.h.s. of Eq.~(\ref{correlator})
is {\em not} evaluated at the on-shell point $K^0 = E_K$. This implies
that for the correlator, in principle, we need to know the off-shell 
behaviour of the emission function, i.e. the quantum mechanical 
structure of the source! Fortunately, nature is nice to us: the
interesting behaviour of the correlator (its deviation from unity) is
concentrated at small values of $\vert \bbox{q} \vert$. Expanding $K^0
= (E_a+E_b)/2$ for small $q$ one finds 
 \begin{equation}
 \label{Konshell}
   K^0 = E_K \, \left( 1 + {\bbox{q}^2 \over 8 E_K^2} + 
   {\cal O}\left({\bbox{q}^4 \over E_K^4}\right) \right) 
   \approx E_K \, .
 \end{equation}
Since the relevant range of $q$ is given by the inverse ``size'' of
the source (to be defined more exactly below), this approximation is
valid as long as the Compton wavelength of the particles is small
compared to this ``source size". For pion, kaon, or proton
interferometry in heavy-ion collisions this is the case, due to the
particle rest masses. This is of enormous practical importance: it
allows one to replace the source Wigner density by a classical
phase-space distribution function for on-shell particles. This
provides a necessary theoretical foundation for the calculation of HBT
correlations from classical hydrodynamic or kinetic (e.g. cascade or
molecular dynamics) simulations of the collision.   

If one approximates the product of single-particle distributions in
the denominator of (\ref{correlator}) by the square of the
spectrum at the average momentum $K$, one obtains the simpler result
 \begin{equation}
 \label{corrapp}
  C(\bbox{q},\bbox{K}) \approx 1 + 
  \left\vert {\int d^4x\, e^{iq{\cdot}x}\, S(x,K) 
              \over
              \int d^4x\, S(x,K)} 
  \right\vert^2
  \equiv 1 + \left\vert \langle e^{iq{\cdot}x} \rangle \right\vert^2
  \, .
 \end{equation}
The deviations from this approximation are proportional to the 
curvature of the single-particle distribution in logarithmic 
representation.\cite{CSH95} They are small in practice because the 
measured single-particle spectra are usually more or less exponential.
In the second equality of (\ref{corrapp}) we defined $\langle \dots 
\rangle$ as the average taken with the emission function; due to the 
$K$-dependence of $S(x,K)$ this average is a function of $K$.
This notation will be used extensively below.

The fundamental relations (\ref{spectrum},\ref{correlator},\ref{corrapp}) 
show that {\em both the single-particle spectrum and the two-particle
  correlation function can be expressed as simple integrals over the
  emission function}. The emission function thus is {\em the} basic
ingredient in the theory of HBT interferometry: once it is 
known, the calculation of one- and two-particle spectra is 
straightforward (even if the evaluation of the integrals may in some 
cases be technically involved). More interestingly, measurements of 
the one- and two-particle spectra provide access to the emission 
function and thus to the space-time structure of the source. This 
latter aspect is, of course, the motivation for exploiting HBT in 
practice. In my talk I will concentrate on the question to what extent
this access to the space-time structure from only momentum-space data
really works, whether it is complete, and (since we will find it is
not and HBT analyses will thus necessarily be model-dependent) what
can be reliably said about the extension and dynamical space-time
structure of the source anyhow, based on a minimal set of intuitive
and highly suggestive model assumptions.  

\subsection{Final state interactions (FSI)}
\label{sec2.3}

Equation (\ref{correlator}),via the plane wave factor $e^{iq{\cdot}x}$
under the integral of the exchange term, reflects the absence of final
state interactions:
 \begin{eqnarray}
   P_2(\bbox{p}_a,\bbox{p}_b) = \int_{x_1,x_2} \!\!\!
   [S\left(x_1,K{+}{\textstyle{q\over2}}\right)\,
    S\left(x_2,K{-}{\textstyle{q\over2}}\right) 
    \pm e^{i q{\cdot}(x_1-x_2)} S(x_1,K) S(x_2,K )] .
 \label{50}
 \nonumber
 \end{eqnarray}
In practice particle interferometry is done with charged particle
pairs which suffer long-range Coulomb final state repulsion on their
way out to the detector. In addition, there may be strong final state
interactions, e.g. in proton-proton interferometry where there is a
strong $s$-wave resonance just above the two-particle threshold. In
this case Eq.~(\ref{50}) must be replaced by~\cite{AHR97}
 \begin{eqnarray}
 \label{50+6}
  &&P_2(\bbox{p}_a,\bbox{p}_b) = \int_{x,y}
       S\left(x+{\textstyle{y\over 2}},p_a\right)\,
       S\left(x-{\textstyle{y\over 2}},p_b\right) 
 \nonumber\\
  &&\qquad \qquad \times \left[ \theta(y^0) 
  \left\vert \phi_{\bbox{q}/2}(\bbox{y}{-}\bbox{v}_b y^0) \right\vert^2
  + \theta(-y^0) 
  \left\vert \phi_{\bbox{q}/2}(\bbox{y}{-}\bbox{v}_a y^0) \right\vert^2
  \right]
 \nonumber\\
  && \pm \int_{x,y} S\left(x+{\textstyle{y\over 2}},K\right)\,
       S\left(x-{\textstyle{y\over 2}},K\right)\,
       \phi^*_{-\bbox{q}/2}(\bbox{y}{-}\bbox{v} y^0) \, 
       \phi_{\bbox{q}/2}(\bbox{y}{-}\bbox{v} y^0) .
 \end{eqnarray}
Here $\bbox{v}{=}\bbox{K}/E_K,\, \bbox{v}_a{=}\bbox{p}_a/E_K,\,
\bbox{v}_b{=}\bbox{p}_b/E_K$ are (to quadratic accuracy in $q$) the
velocities of the particles with momentum $\bbox{K},\,\bbox{p}_a,\,
\bbox{p}_b$, respectively, and $\phi_{\bbox{q}/2}(\bbox{r})$
is an FSI distorted wave with asymptotic relative momentum
$\bbox{q}/2$, evaluated at the two-particle relative distance
$\bbox{r}$. Upon replacing the latter by plane waves (\ref{50+6}) turns
into (\ref{50}). The FSI distorted waves can be calculated by solving
a non-relativistic Schr\"odinger equation for the relative motion
which includes the FSI potential {\em in the rest system of the pair}
(where $\bbox{K} = \bbox{v} =0$). Eq.~(\ref{50+6}) represents a
non-relativistic Galilei-transformation of the result from the pair
rest frame to the frame in which $\bbox{p}_a$ and $\bbox{p}_b$ are
measured; therefore it is only valid in observer frames in which the
pair moves non-relativistically. In order to evaluate Eq.~(\ref{50+6})
one must therefore first transform the 4-momenta $p_{a,b}$ to such a
frame (best directly into the pair rest frame). The momentum argument
$\bbox{q}$ of the FSI distorted waves $\phi$ is then the difference
between the two spatial momenta {\em in that frame}, and their
space-time argument $\bbox{y} - \bbox{v}_iy^0$ is the relative
distance of the two particles in that frame {\em at the time when the
  second particle is emitted}.\cite{AHR97} Since the latter depends
not only on the time difference $y^0$ between emission points, but
also on the velocity of the first emitted particle, these arguments
depend on the momentum argument of the emission function associated
with the first emitted particle. The two terms $\sim \theta(\pm y^0)$
in the direct term reflect the two possible time orderings between the
emission points.

Equations~(\ref{50}) and (\ref{50+6}) can be implemented into event
generators, following the procedure given elsewhere.\cite{AHR97,zhang97} 

\subsection{The redundance of wavepackets}
\label{sec2.4}

In the last two years it has been repeatedly suggested that, due to
the smallness of the sources, the theory of particle interferometry in
high energy physics should be based on a finite-size wave-packet
description rather than on plane wave propagation from the source to
the detector.\cite{zhang97,Wetal,ZC98,W98,MP98} I will argue that, if
done correctly, it doesn't matter whether you start from wave packets
or not, and that the free parameter $\sigma$ (the wave packet width)
occurring in these approaches is essentially unmeasurable (except,
perhaps, in elementary $e^+e^-$ or $pp$ collisions).

To the extent that the detector really measures the momenta of the
particles (which it is supposed to do with highest possible accuracy),
the measurement process can be described as a projection at $t=\infty$
of the emitted pion state on a plane wave momentum eigenstate,
irrespective of the actual localization properties of the emitted
states; the latter will be reflected in the momentum spectrum
resulting from this projection. The usefulness of HBT interferometry
rests exactly on the fact that free propagation after emission from
the source does not change this overlap integral, i.e. that the 1- and
2-particle momentum spectra remain unchanged on their way from source
to detector. Otherwise data measured meters away from the
collision could not be used to extract information about the reaction
zone. Final state interactions spoil this feature; therefore they must
be accounted for analytically or numerically via Eq.~(\ref{50+6})
before source information can be extracted from the measured
correlator. For free propagation, however, the spectrum calculated
from the overlap between the emitted wavefunction and the 
momentum eigenstates is identical if calculated at the time of
particle freeze-out or at the detector time several nanoseconds
later.\cite{AHR97} In both cases, the relation between the spectra and
the source distribution $S(x,K)$ is given by Eqs.~(\ref{spectrum}) and
(\ref{correlator}). 

While wave packets are a physically intuitive concept, one must take
care not to exaggerate their role in particle interferometry. I will
show that, to first order, their intrinsic structure can be completely
absorbed into the emission function, leaving no measurable trace of
the wave packet width. Statements to the contrary~\cite{ZC98,MP98} are
at best misleading. To prove my point, let me anticipate from 
Sec.~\ref{sec3.1} that the primary feature of the two-particle
correlator, its Gaussian width in $\bbox{q}$, determines only the
r.m.s. width of the source in space-time; the extraction of finer
structures of the source requires a quantitative study of the
deviations from Gaussian behaviour of the correlator and therefore at
least an order of magnitude more accurate data. The dominant features
of the correlator can thus be reproduced by replacing the true source
function $S(x,K)$ by a Gaussian with the same r.m.s. width in space-time. 

Let us now follow custom~\cite{zhang97,Wetal,ZC98,W98,MP98} and assume
that the source emits at times $\tau_i$ Gaussian wave packets of
spatial width $\sigma$, centered at points $\bbox{\xi}_i$ in
coordinate space and $\bbox{\pi}_i$ in momentum space:
 \begin{equation}
 \label{wavepacket}
   \psi_i(\bbox{x},\tau_i) = {1\over (\pi\sigma^2)^{3/4}}
   \, \exp\left(-{(\bbox{x}-\bbox{\xi}_i)^2\over 2\sigma^2} +
     i\bbox{\pi}_i\cdot \bbox{x}\right)\, .
 \end{equation}
The Wigner density corresponding to such an individual wave packet is
given by
 \begin{equation}
 \label{Wigwave}
   S_{\rm w.p.}(x,\bbox{p}) = {E_p\over \pi^3} \delta(x^0-\tau_i) \, 
   \, \exp\left(-{(\bbox{x}-\bbox{\xi}_i)^2\over \sigma^2} 
         -\sigma^2 (\bbox{p}-\bbox{\pi}_i)^2\right)\, ;
 \end{equation}
it is normalized to $\int (d^3p/E_p) \int d^4x \, S_{\rm
  w.p.}(x,\bbox{p}) = 1$. Its r.m.s. width parameters $\Delta x =
\sigma/\sqrt{2}$ and $\Delta p = 1/(\sqrt{2}\sigma$ in the three
Cartesian directions saturate the uncertainty relation $\Delta x\Delta
p \geq \hbar/2$.  

Let us distribute the wave packet emission points $(\tau_i,
\bbox{\xi}_i, \bbox{\pi}_i)$ according to the {\em classical} Gaussian
phase-space distribution 
 \begin{equation}
 \label{rhogauss}
   \rho(\tau,\bbox{\xi},\bbox{\pi}) =  \langle N\rangle 
   {\exp\left(- {(\tau-\tau_0)^2\over 2(\Delta\tau)^2}\right) 
    \over \sqrt{2\pi(\Delta\tau)^2}}\,
   {\exp\left( - {\bbox{\xi}^2\over 2 R_0^2} 
               - {\bbox{\pi}^2\over 2 \Delta_0^2}\right)
    \over (2\pi R_0\Delta_0)^3}\, ,
 \end{equation}
normalized to the average pion multiplicity $\langle N \rangle$ per event.
The r.m.s. widths $\Delta\tau, R_0, \Delta_0$ of this classical
probability density are not constrained by quantum mechanics. It can
be shown \cite{zhang97,Wetal,ZC98,W98} that the spectra and
correlation functions are then given by Eqs.~(\ref{spectrum}) and
(\ref{correlator}) with an emission function $S(x,K)$ which is
obtained by folding the classical phase-space distribution
(\ref{rhogauss}) with the intrinsic Wigner density (\ref{Wigwave}) of
the wave packets:  
 \begin{eqnarray}
 \label{folding}
    S(x,\bbox{K}) &=& \int d\tau\, d^3\xi\, d^3\pi \,
    \rho(\tau,\bbox{\xi},\bbox{\pi}) \, 
    S_{\rm w.p.}(x^0-\tau,\bbox{x}-\bbox{\xi},\bbox{K}-\bbox{\pi})
 \nonumber\\
    &=& \langle N\rangle 
        {\exp\left(- {(x^0-\tau_0)^2\over 2(\Delta\tau)^2} \right) 
         \over \sqrt{2\pi(\Delta\tau)^2}}\,
      {\exp\left( - {\bbox{x}^2\over 2 R^2} 
                  - {\bbox{K}^2\over 2 \Delta^2}\right)
       \over (2\pi R\Delta)^3}\, ,
 \\
 \label{RDelta}
    R^2 &=& R_0^2 + {\sigma^2\over 2}\, , \qquad
    \Delta^2 = \Delta_0^2 + {1\over 2 \sigma^2}\, .
 \end{eqnarray}
Only the combinations $R,\Delta$ from (\ref{RDelta}) enter the 1- and
2-particle spectra; at the Gaussian level there exists no measurement
which allows to disentangle $R_0, \Delta_0$ and $\sigma$. As long as
$R\Delta \gg \hbar/2$, there are infinitely many combinations of
$R_0,\Delta_0,\sigma$ which describe the same source. Only if $R$ and
$\Delta$ saturate the uncertainty limit, $R\Delta = \hbar/2$, one may
conclude $R_0=\Delta_0=0$ and $\sigma = \sqrt{2}R$.

{\em The wave packet size $\sigma$ is thus generally not measurable};
as long as the effective emission function $S(x,\bbox{K})$ (i.e. $R,
\Delta$) is not changed, variations of $\sigma$ have no influence on
the momentum distributions. Final state interactions do not affect
this reasoning either -- instead of (\ref{correlator}) one must then
evaluate (\ref{50+6}) with the same emission function
(\ref{folding}). The effect of 2-particle final state interactions on
the time evolution of the wave packets~\cite{MP98} is fully taken into
account by weighting the effective emission function $S(x,K)$ with the
distorted waves $\phi_{\bbox{q}/2}$ in (\ref{50+6}); a generalization
which includes also the FSI with the electric charge of the central
fireball is known \cite{Barz,AH98}. There is no need to describe the
time evolution of the wave packets explicitly as done by Merlitz and 
Pelte~\cite{MP98} using a sophisticated numerical algorithm. Finally,
the above statement remains true~\cite{HW98} even if multi-boson
symmetrization effects~\cite{ZC98,W98} are included. 

I have presented the argument using Gaussian wave packets and a
Gaussian parametrization for the distribution of their centers.
As explained above this is sufficient since, in leading order, HBT
interferometry gives only access to the r.m.s. widths of the source
such that one cannot distinguish Gaussians from other source
shapes. At the next level of accuracy these statements are surely
subject to (small) corrections. So far, however, I cannot see how in
heavy-ion collisions the wave packet size might be determined
experimentally. For this reason I prefer avoiding the concept of wave
packets, thereby eliminating the poorly controlled free parameter
$\sigma$ from the theory of HBT interferometry in nuclear physics. 

From here on I will neglect final state interactions, assuming that
the measured correlators have been or can be corrected for them.

\subsection{The mass-shell constraint}
\label{sec2.5}

Expressions (\ref{correlator},\ref{corrapp}) show that the correlation 
function is related to the emission function by a Fourier 
transformation. At first sight this might suggest that one should 
easily be able to reconstruct the emission function from the measured 
correlation function by inverse Fourier transformation, the single 
particle spectrum (\ref{spectrum}) providing the normalization. This 
is, however, not correct. The reason is that, since the correlation 
function is constructed from the on-shell momenta of the measured 
particle pairs, not all four components of the relative momentum $q$ 
occurring on the r.h.s. of (\ref{corrapp}) are independent. They are 
related by the ``mass-shell constraint" (\ref{ortho}) which can, for 
instance, be solved for $q^0$:
 \begin{equation}
 \label{massshell}
   q^0 = \bbox{\beta}\cdot \bbox{q} \qquad {\rm with} \qquad 
   \bbox{\beta} = {\bbox{K}\over K^0} \approx {\bbox{K}\over E_K}\, .
 \end{equation} 
$\bbox{\beta}$ is (approximately) the velocity of the c.m. of the 
particle pair. The Fourier transform in (\ref{corrapp}) is therefore 
not invertible, and the reconstruction of the space-time structure of 
the source from HBT measurements will thus always require additional 
model assumptions. 

It is instructive to insert (\ref{massshell}) into (\ref{corrapp}): 
 \begin{equation}
 \label{corrapp1}
  C(\bbox{q},\bbox{K}) \approx 1 + 
  \left\vert {\int d^4x\, \exp\bigl( 
                          i\bbox{q}{\cdot}(\bbox{x}-\bbox{\beta}\, t) 
                          \bigr) \, S(x,K)
              \over
              \int d^4x\, S(x,K)} 
  \right\vert^2 \, .
 \end{equation}
This shows that the correlator $C(\bbox{q},\bbox{K})$ actually mixes the 
spatial and temporal information on the source in a non-trivial way 
which depends on the pair velocity $\bbox{\beta}$. Only for a
pulsed source $\sim \delta(t-t_0)$ things are simple: the correlator then 
just measures the Fourier transform of the spatial source 
distribution. 

It is instructive to look at the problem also in the following way:
If one rewrites Eq.~(\ref{corrapp1}) in the pair rest frame where
$\bbox{K}=0$ and hence $q^0=0$, one obtains 
 \begin{equation}
 \label{restframe}
   C(\bbox{q,K})- 1 = \int d^3r\, \cos(\bbox{q\cdot r})\,
   S_{\rm rel}(\bbox{r};K)
 \end{equation}
where
 \begin{equation}
 \label{reldis}
   S_{\rm rel}(\bbox{r};K) = \int d^3R\  
   \bar s_K(\bbox{R} + {\textstyle{1\over 2}} \bbox{r})\ 
   \bar s_K(\bbox{R} - {\textstyle{1\over 2}} \bbox{r})\, ,
 \end{equation}
with
 \begin{equation}
 \label{sbar}
   \bar s_K(\bbox{x}) = \int dt\ s(\bbox{x},t; K) 
   = \int dt\ {S(x,K) \over \int d^4x'\, S(x',K)}\, ,
 \end{equation}
is the {\em time-integrated} normalized relative distance distribution
in the source. The latter can, in principle, be uniquely reconstructed
from the measured correlator~\cite{brown97} by inverting the
cosine-Fourier transform (\ref{restframe}). But since it gives only
the time integral of the relative distance distribution for fixed pair
momentum $\bbox{K}$ in the pair rest frame, no direct information on
the time structure of the source is obtainable! Only by looking at the
result as a function of $\bbox{K}$, which, as I will show, brings out
the collective dynamical features of the source, can one hope to
unfold the time-dependence of the emission function. It is clear that
this will be only possible within the context of specific source
parametrizations. 

\subsection{$K$-dependence of the correlator}
\label{sec2.6}

We have seen that in general the correlator is a function of {\em
  both} $\bbox{q}$ and $\bbox{K}$. Only if the emission function  
factorizes in $x$ and $K$, $S(x,K) = F(x)\,G(K)$ (no
``$x$-$K$-correlations'', i.e. every point $x$ in the source emits
particles with the same momentum spectrum $G(K)$), the $K$-dependence in 
$G(K)$ cancels between numerator and denominator of (\ref{corrapp}), 
and the correlator seems to be $K$-independent. However, not even 
this is really true: even after the cancellation of the explicit 
$K$-dependence $G(K)$, there remains an implicit $K$-dependence
via the pair velocity $\bbox{\beta} \approx \bbox{K}/E_K$ 
in the exponent on the r.h.s. of Eq.~(\ref{corrapp1})! Only if both 
conditions, factorization of the emission function in $x$ and $K$ {\em 
and} instantaneous emission $\sim \delta(t-t_0)$, apply
simultaneously, the correlation function is truely $K$-independent.
Even for stars which are time-independent sources there remains a
$K$-dependence: the correlator is then~\cite{He97} $\sim
\delta(\bbox{\beta}\cdot\bbox{q})$, i.e. there are only transverse
($\perp\bbox{\beta}$), but no longitudinal ($\parallel\bbox{\beta}$)
correlations. 

It is hard to believe that this complication in the application of the
original HBT idea to high-energy collisions went nearly unnoticed for
more than 20 years. It was only brought to light in 1984 by Scott
Pratt~\cite{P84} in his pioneering work on HBT interferometry for
heavy-ion collisions.   

If one parametrises the correlator by a Gaussian in $q$ (see below)
this means that in general the parameters (``HBT radii'') depend on
$K$. Typical sources of $x$-$K$ correlations in the emission function
are a collective expansion of the emitter and/or temperature gradients in 
the particle source: in both cases the momentum spectrum $\sim \exp[-
p{\cdot}u(x)/T(x)]$ of the emitted particles (where $u^\mu(x)$ is the 
4-velocity of the expansion flow) depends on the emission point. In 
the case of collective expansion, the spectra from different emission 
points are Doppler shifted relative to each other. If there are
temperature gradients, e.g. a high temperature in the center and 
cooler matter at the edges, the source will look smaller for 
high-momentum particles (which come mostly from the hot center) than 
for low-momentum ones (which receive larger contributions also from 
the cooler outward regions).

We thus see that collective expansion of the source induces a 
$K$-dependence of the correlation function. But so do temperature 
gradients. The crucial question is: does a careful measurement of the 
correlation function, in particular of its $K$-dependence, permit a 
separation of such effects, i.e. can the collective dynamics of the 
source be quantitatively determined through HBT experiments? We will 
see that this is not an easy task; however, with sufficiently good 
data, it should be possible. In any case, the $K$-dependence of the 
correlator is a decisive feature which puts the HBT game into a 
completely new ball park. Two-particle correlation measurements which
are not able to resolve the $K$-dependence of the HBT parameters are,
in high energy nuclear and particle physics, only of very limited use.

\section{Model-independent discussion of HBT correlation functions}
\label{sec3}
\subsection{The Gaussian approximation -- HBT radii as homogeneity
  lengths} 
\label{sec3.1} 

The most interesting feature of the two-particle correlation function
is its half-width. Actually, since the relative momentum $\bbox{q} =
\bbox{p}_1 - \bbox{p}_2$ has three Cartesian components, the fall-off
of the correlator for increasing $q$ is not described by a single
half-width, but rather by a (symmetric) 3$\times$3 tensor~\cite{CNH95}
which describes the curvature of the correlation function near
$\bbox{q} = 0$. We will see that in fact nearly all relevant
information that can be extracted from the correlation function
resides in the 6 independent components of this tensor. This in turn
implies that in order to compute the correlation function $C$ it is
sufficient to approximate the source function $S$ by a Gaussian in $x$
which contains only information on its space-time moments up to second
order.  

Let us write the arbitrary emission function $S(x,K)$ in the following
form: 
 \begin{equation}
 \label{7}
   S(x,K) = N(K)\  S(\bar x(K),K)\ 
            e^{ - \half \tilde x^\mu(K)\, B_{\mu\nu}(K)\, \tilde x^\nu(K)}
   + \delta S(x,K) \, ,
 \end{equation} 
where we adjust the parameters $N(K)$, $\bar x^\mu(K)$, and 
$B_{\mu\nu}(K)$ of the Gaussian first term in such a way that the 
correction term $\delta S$ has vanishing zeroth, first and second 
order space-time moments:
 \begin{equation}
 \label{deltaS}
   \int d^4x\, \delta S(x,K) = 
   \int d^4x\, x^\mu\, \delta S(x,K) = 
   \int d^4x\, x^\mu x^\nu\,  \delta S(x,K) = 0\, .
 \end{equation} 
This is achieved by setting
 \begin{eqnarray}
  N(K) &=& E_K {dN\over d^3 K}\,
           {\det^{1/2} B_{\mu\nu}(K) \over S(\bar x(K),K)}\, ,
 \label{NK}\\
  \bar x^\mu(K) &=& \langle x^\mu \rangle\, , 
 \label{barx}\\
  \left(B^{-1}\right)_{\mu\nu}(K) 
  &=& \langle \tilde x_\mu \tilde x_\nu \rangle 
      \equiv \langle (x -\bar x)_\mu (x- \bar x)_\nu \rangle \, .
 \label{Bmunu}
 \end{eqnarray}
The ($K$-dependent) average over the source function $\langle \dots 
\rangle$ has been defined in Eq.~(\ref{corrapp}). The normalization factor 
(\ref{NK}) ensures that the Gaussian term in (\ref{7}) gives the 
correct single-particle spectrum (\ref{spectrum}); it fixes the 
normalization on-shell, i.e. for $K^0=E_K$, but as we discussed this 
is where we need the emission function also for the computation of the 
correlator. $\bar x(K)$ in (\ref{barx}) is the centre of the emission
function $S(x,K)$ and approximately equal to its ``saddle point",
i.e. the point of highest emissivity for particles with momentum $K$. The 
second equality in (\ref{Bmunu}) defines $\tilde x$ as the space-time 
coordinate relative to the centre of the emission function; only this 
quantity enters the further discussion, since, due to the invariance 
of the momentum spectra under arbitrary translations of the source in 
coordinate space, the absolute position of the emission point is not 
measurable in experiments which determine only particle momenta. Since 
$\bar x(K)$ is not measurable, neither is the normalization
$N(K)$~\cite{HTWW96} as its definition (\ref{NK}) involves the
emission function at $\bar x(K)$. Finally, Eq.~(\ref{Bmunu}) ensures that the 
Gaussian first term in (\ref{7}) correctly reproduces the second
central space-time moments $\langle \tilde x_\mu \tilde x_\nu \rangle$
of the original emission function, in particular its r.m.s. widths in
the various space-time directions.  

Inserting the decomposition (\ref{7}) into Eq.~(\ref{corrapp}) we 
obtain
 \begin{equation}
 \label{corrgauss}
   C(\bbox{q},\bbox{K}) = 1 + \exp\bigl[
   - q^\mu q^\nu \langle \tilde x_\mu \tilde x_\nu \rangle(\bbox{K}) 
   \bigr]
   + \delta C(\bbox{q},\bbox{K})\, .
 \end{equation}
The Gaussian in $q$ results from the Fourier transform of the Gaussian 
contribution in (\ref{7}); the last term $\delta C$ receives 
contributions from the second term $\delta S$ in (\ref{7}) which 
contains information on the third and higher order space-time moments 
of the emission function, like sharp edges, wiggles, secondary peaks,
or non-Gaussian tails in the source. It is at least of order $q^4$; 
the second derivative of the full correlator at $q=0$ is
given {\em exactly} by the Gaussian in (\ref{corrgauss}). 

In the past it has repeatedly been observed that the correlation data 
appear to be better fit by exponentials than by Gaussians. As far as I
know, however, this happened usually for 1-dimensional fits as a
function of the single Lorentz invariant variable $Q_{\rm inv}^2 =
(q^0)^2 -\bbox{q}^2$ while, at least for heavy ion collisions, the
3-dimensional correlators look much more Gaussian. (Correlators from
elementary collisions are more strongly affected by resonance 
decay contributions (Sec.\ref{sec6}), giving rise to severe deviations 
from Gaussian behavior even in three $q$-dimensions.\cite{EGHW98}) 
Contemplating the structure of Eq.~(\ref{corrgauss}) one realizes that
a fit in $Q_{\rm inv}^2$ does not make sense: the generic structure of
the exponent, $-q^\mu q^\nu \langle \tilde x_\mu \tilde x_\nu
\rangle$, tells us that the term $(q^0)^2$ should come with the time
variance of the source while the spatial components $(q^i)^2$ should
come with the spatial variances of the source. Since all variances are
positive semidefinite by definition, it does not make sense to
parametrize the correlation function by a variable in which $(q^0)^2$
and $\bbox{q}^2$ appear with the opposite sign! Such a fit might
work if the time variance and all mixed variances vanished
identically and all three spatial variances were equal, but this is
certainly not generic and also not frame-independent. {\em The
  variable $Q_{\rm inv}$ should therefore {\em not} be used for 
  fitting HBT data.} 

Note that Eq.~(\ref{corrgauss}) has no factor $\half$ in 
the exponent. If the measured correlator is fitted by a
Gaussian as defined in (\ref{corrgauss}), its $q$-width 
can be directly interpreted in terms of the r.m.s. widths of the 
source in coordinate space. Model comparisons are thus most easy
if the latter are directly parametrized in terms of r.m.s. widths.

Eqs.~(\ref{7}) and (\ref{corrgauss}) would not be useful 
if the contributions from $\delta S$ and $\delta C$ were not somehow 
small enough to be neglected. This requires a numerical investigation.
It was shown~\cite{WSH96} that in typical (and even in some not so
typical) situations $\delta S$ has a negligible influence on the half
width of the correlation function. It contributes only weak, 
essentially unmeasurable structures in $C(\bbox{q},\bbox{K})$ at large
values of $\bbox{q}$. The reader can easily verify this analytically
for an emission function with a sharp box profile; the results for the
exact correlator and the one resulting from the Gaussian approximation
(\ref{7}) differ by less than 5\% in the half widths;\cite{CNH95} the
exact correlator has, as a function of $q$, secondary maxima with an
amplitude below 5\% of the value of the correlator at $q=0$. We have
checked that similar statements remain even true for a source with a
doughnut structure, i.e. with a hole in the middle, which was obtained
by rotating the superposition of two 1-dimensional Gaussians separated
by twice their r.m.s. widths around their center. The only
situation where these statements require qualification is if the
correlator receives contributions from the decay of long-lived
resonances; unfortunately, this is of relevance for pion
interferometry as will be discussed in Sec.~\ref{sec6}. 

Eq.~(\ref{corrgauss}) implies that the two-particle correlator
measures the second central space-time moments of the emission
function. That's it -- finer features of its space-time structure
(edges, wiggles, holes) cannot be measured with two-particle
correlations, but require the analysis of three-, four-, \dots,
many-particle correlations.\cite{HZ97} It follows that 
possible rapid quantum oscillations of the source Wigner density 
are also essentially unmeasurable by 2-particle interferometry. The
variances $\langle \tilde x_\mu \tilde x_\nu \rangle$ are in general
{\em not} identical with our naive intuitive notion of the ``source
radius": unless the source is stationary and has no
$x$-$K$-correlations at all, the variances depend on the pair momentum
$\bbox{K}$ and cannot be interpreted in terms of simple overall source
geometry. Their correct interpretation~\cite{CSH95,MS88,AS95} is in
terms of ``lengths of homogeneity" which give, for each pair momentum
$\bbox{K}$, the size of the region around the point of maximal
emissivity $\bar x(\bbox{K})$ over which the emission function is
sufficiently homogeneous to contribute to the correlator. Thus HBT 
measures ``regions of homogeneity" in the source and their variation
with the momentum of the particle pairs. As we will see, the latter is
the key to their physical interpretation.   

\subsection{YKP parametrization for the correlator and HBT radius
  parameters}
\label{sec3.2}

A full characterization of the source in terms of its second order 
space-time variances requires knowledge of the 10 functions $\langle
\tilde x_\mu \tilde x_\nu \rangle(\bbox{K})$. However, due to the
mass-shell constraint (\ref{massshell}) which leaves only three
independent components of $q$, only 6 linear combinations of theses
functions are actually measurable.\cite{CNH95} For azimuthally
symmetric sources three of these 10 functions vanish by
symmetry~\cite{CNH95}, but again the mass-shell constraint permits 
to measure only 4 linear combinations of the remaining 7 functions
of $\bbox{K}$.

Before the correlator (\ref{corrgauss}) can thus be fit to data, the
redundant components of $q$ must first be eliminated via
(\ref{massshell}). We use a Cartesian coordinate system with the
$z$-axis along the beam direction and the $x$-axis along
$\bbox{K}_\perp$. Then $\bbox{\beta} = (\beta_\perp, 0, \beta_l)$. We
assume an azimuthally symmetric source (impact parameter $\approx 0$)
and eliminate from (\ref{corrgauss}) $q_x$ and $q_y$ in terms of
$q_\perp^2 = q_x^2 + q_y^2$, $q_l$ and $q^0$. This yields the YKP
parametrization\cite{CNH95,HTWW96} 
 \begin{equation}
 \label{18}
   C(\bbox{q,K}) = 1 +  
     \exp\biggl[ - R_\perp^2 q_\perp^2 
                 - R_\parallel^2 \left( q_l^2 - (q^0)^2 \right) 
                 - \left( R_0^2 + R_\parallel^2 \right)
                         \left(q\cdot U\right)^2
                \biggr] .
 \end{equation}
Here $R_\perp$, $R_\parallel$, $R_0$, $U$ are four $K$-dependent
parameter functions. $U(\bbox{K})$ is a 4-velocity with only a
longitudinal spatial component:  
 \begin{equation}
 \label{19}
   U(\bbox{K}) = \gamma(\bbox{K}) \left(1, 0, 0, v(\bbox{K}) \right) ,
   \ \ {\rm with} \ \
   \gamma = {1\over \sqrt{1 - v^2}}\, .
 \end{equation}
Its value depends, of course, on the measurement frame. The ``Yano-Koonin 
velocity'' $v(\bbox{K})$ can be calculated~\cite{HTWW96} in an
arbitrary reference frame from the second central space-time moments
of $S(x,K)$. It is, to a good approximation, the longitudinal velocity
of the fluid element from which most of the particles with momentum
$\bbox{K}$ are emitted.\cite{CNH95,HTWW96} For sources with
boost-invariant longitudinal expansion velocity the YK-rapidity
associated with $v(\bbox{K})$ is linearly related to the pair rapidity
$Y$.\cite{HTWW96}

The other three YKP parameters do not depend on the longitudinal 
velocity of the observer. (This distinguishes the YKP form (\ref{18})
from the Pratt-Bertsch parametrization~\cite{HB95,CSH95,P84} which
results from eliminating $q^0$ in (\ref{corrgauss}).) Their physical
interpretation is easiest in terms of coordinates measured in the
frame where $v(\bbox{K})$ vanishes. There they are given by~\cite{CNH95} 
 \begin{eqnarray}   
   R_\perp^2(\bbox{K}) &=& \langle \tilde y^2 \rangle \, ,
 \label{20a} \\
   R_\parallel^2(\bbox{K}) &=& 
   \left\langle \left( \tilde z - (\beta_l/\beta_\perp) \tilde x
                \right)^2 \right \rangle   
     - (\beta_l/\beta_\perp)^2 \langle \tilde y^2 \rangle 
     \approx \langle \tilde z^2 \rangle \, ,
 \label{20b} \\
   R_0^2(\bbox{K}) &=& 
   \left\langle \left( \tilde t -  \tilde x/\beta_\perp
                \right)^2 \right \rangle 
    -  \langle \tilde y^2 \rangle/\beta_\perp^2 
    \approx \langle \tilde t^2 \rangle .
 \label{20c}
 \end{eqnarray}
$R_\perp$, $R_\parallel$ and $R_0$ thus measure, approximately, the 
($K$-dependent) transverse, longitudinal and temporal regions of 
homogeneity of the source in the local comoving frame of the emitter. 
The approximation in (\ref{20b},\ref{20c}) consists of dropping terms 
which for the model discussed below vanish in the absence of
transverse flow and were found to be small even for finite transverse
flow.\cite{HTWW96} Note that it leads to a complete separation of 
the spatial and temporal aspects of the source. This separation is
spoiled by sources with $\langle \tilde x^2 \rangle \ne \langle \tilde
y^2 \rangle$. For our source this happens for non-zero transverse (in
particular for large) transverse flow $\eta_f$, but for opaque sources
where particle emission is surface dominated~\cite{HV96} this occurs
even without transverse flow.\cite{HV96,TH98b}

\section{A model for a finite expanding source}
\label{sec4}

For our quantitative studies we used the following model for an
expanding thermalized source:~\cite{CNH95}  
 \begin{equation}
 \label{3.15}
   \!\!\! S(x,K)\! =\! 
   {M_\perp \cosh(\eta{-}Y) \over 8 \pi^4 \Delta \tau}
    \exp\!\! \left[-{K{\cdot}u(x) \over T(x)} 
                   - {(\tau{-}\tau_0)^2 \over 2(\Delta \tau)^2}
                   - {r^2 \over 2 R^2}
                   - {(\eta{-}\eta_0)^2 \over 2 (\Delta \eta)^2}\right].
 \end{equation}
Here $r^2 = x^2+y^2$, the spacetime rapidity $\eta = {1 \over 2} 
\ln[(t+z)/(t-z)]$, and the longitudinal proper time $\tau= \sqrt{t^2-
z^2}$ parametrize the spacetime coordinates $x^\mu$, with measure 
$d^4x = \tau\, d\tau\, d\eta\, r\, dr\, d\phi$. $Y = {1\over 2} 
\ln[(E_K+K_L)/(E_K-K_L)]$ and $M_\perp = \sqrt{m^2 + K_\perp^2}$ 
parametrize the longitudinal and transverse components of the pair 
momentum $\bbox{K}$. $\sqrt{2} R$ is the transverse geometric
(Gaussian) radius of the source, $\tau_0$ its average freeze-out
proper time, $\Delta \tau$ the mean proper time duration of particle
emission, and $\Delta \eta$ parametrizes the finite longitudinal
extension of the source. $T(x)$ is the freeze-out temperature; if you
don't like the idea of thermalization in heavy ion collisions, you can
think of it as a parameter that describes the random distribution of
the particle momenta at each space-time point around their average
value. The latter is parametrized by a collective flow velocity
$u^\mu(x)$ in the form 
 \begin{equation}
 \label{26}
   u^\mu(x) = \left( \cosh \eta \cosh \eta_t(r), \,
                     \sinh \eta_t(r)\, \bbox{e}_r,  \,
                     \sinh \eta \cosh \eta_t(r) \right) ,
 \end{equation}
with a boost-invariant longitudinal flow rapidity $\eta_l = \eta$ 
($v_l = z/t$) and a linear transverse flow rapidity profile 
 \begin{equation}
 \label{27}
  \eta_t(r) = \eta_f \left( {r \over R} \right)\, .
 \end{equation} 
$\eta_f$ scales the strength of the transverse flow. The exponent of 
the Boltzmann factor in (\ref{3.15}) can then be written as
 \begin{equation}
 \label{26a}
  K\cdot u(x) = M_\perp \cosh(Y-\eta) \cosh\eta_t(r) - 
                \bbox{K}_\perp{\cdot}\bbox{e}_r \sinh\eta_t(r)\, .
 \end{equation}
For vanishing transverse flow ($\eta_f=0$) the source depends only 
on $M_\perp$, and remains azimuthally symmetric for all $K_\perp$.                                                 
Since in the absence of transverse flow the $\beta$-dependent terms in 
(\ref{20b}) and (\ref{20c}) vanish and the source itself depends only 
on $M_\perp$, all three YKP radius parameters then show perfect 
$M_\perp$-scaling. Plotted as functions of $M_\perp$, they coincide 
for pion and kaon pairs (see Fig.~\ref{F1}, left column). For non-zero
transverse flow (right column) this $M_\perp$-scaling is broken by two
effects: (1) The thermal exponent (\ref{26a}) receives an additional
contribution proportional to $K_\perp = \sqrt{ M_\perp^2 - m^2}$. (2)
The terms which were neglected in the second equalities of
(\ref{20b},\ref{20c}) are non-zero, and they also depend on
$\beta_\perp = K_\perp/E_K$. Both effects induce an explicit rest mass
dependence and destroy the $M_\perp$-scaling of the YKP size parameters.  

\section{$\bbox{K}$-dependence of YKP parameters and collective flow}
\label{sec5}

Collective expansion induces correlations between coordinates and 
momenta in the source, and these result in a dependence of the HBT 
parameters on the pair momentum $K$. At each point in the source the 
local velocity distribution is centered around the average fluid 
velocity; two points whose fluid elements move rapidly relative to 
each other are thus unlikely to contribute particles with small 
relative momenta. Essentially only such regions in the source 
contribute to the correlation function whose fluid elements move with 
velocities close to the velocity of the observed particle pair. 

\subsection{The Yano-Koonin velocity and longitudinal flow}
\label{sec5.1}

\begin{figure}[h]
\vspace*{4.3cm}
\includegraphics{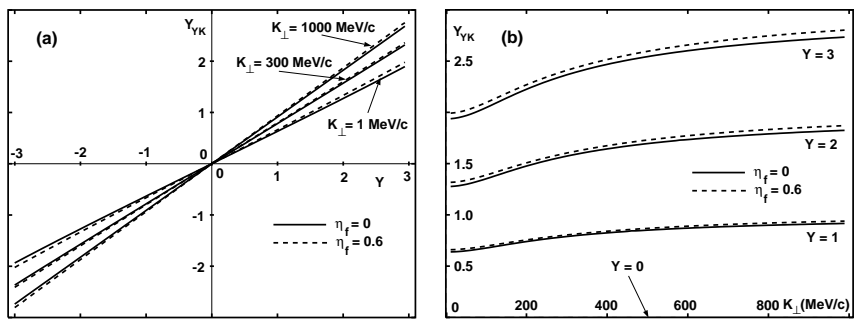}
\caption{(a) The Yano-Koonin rapidity for pion pairs, as a function of
  the pair c.m. rapidity $Y$, for various values of $K_\perp$ and two
  values for the transverse flow $\eta_f$. (b) The same, but plotted
  against $K_\perp$ for various values of $Y$ and $\eta_f$. Source
  parameters: $T=140$ MeV, $\Delta\eta=1.2$, $R=3$ fm, $\tau_0=3$
  fm/$c$, $\Delta\tau=1$ fm/$c$. 
\label{F1}} 
\end{figure}

Fig.~\ref{F1} shows (for pion pairs) the dependence of the YK velocity
on the pair momentum $\bbox{K}$. In Fig.~\ref{F1}a we show the YK
rapidity $Y_{_{\rm YK}} = \frac 12 \ln[(1+v)/(1-v)]$ as a function of
the pair rapidity $Y$ (both relative to the CMS) for different values
of $K_\perp$, in Fig.~\ref{F1}b the same quantity as a function of
$K_\perp$ for different $Y$.  Solid lines are without transverse flow,
dashed lines are for $\eta_f=0.6$.  For large $K_\perp$ pairs, the YK
rest frame approaches the LCMS (which moves with the pair rapidity
$Y$); in this limit all pairs are thus emitted from a small region in
the source which moves with the same longitudinal velocity as the
pair. For small $K_\perp$ the YK frame is considerably slower than the
LCMS; this is due to the thermal smearing of the particle velocities
in our source around the local fluid velocity $u^\mu(x)$.\cite{HTWW96}
The linear relationship between the rapidity $Y_{_{\rm YK}}$ of the
Yano-Koonin frame and the pion pair rapidity $Y$ is a direct
reflection of the boost-invariant longitudinal expansion
flow.\cite{HTWW96} For a non-expanding source $Y_{_{\rm YK}}$ would be
independent of $Y$. Additional transverse flow is seen to have nearly
no effect. The dependence of the YK velocity on the pair rapidity thus
measures directly the longitudinal expansion of the source and cleanly
separates it from its transverse dynamics. 

The NA49 data for 160 A GeV Pb+Pb collisions~\cite{NA49} show
very clearly such a more or less linear rise of the Yano-Koonin source
rapidity with the rapidity of the pion pair. This confirms, in the
most transparent way imaginable, their earlier
conclusion~\cite{alber95} based on the $Y$-dependence of the 
longitudinal radius pa\-ra\-me\-ter $R_l$ in the Pratt-Bertsch
parametrization that the source created in 200 A GeV S+$A$ collisions
expands longitudinally in a nearly boost-invariant way.

Note that this longitudinal flow need not be of hydrodynamical
(pressure generated) nature. Similar longitudinal position-momentum
correlations arise in string fragmentation. This should cause a
similar linear rise of the YK-rapidity with the pair rapidity in jet
fragmentation (with the $z$-axis oriented along the jet axis). It
would be interesting to confirm this prediction in $e^+e^-$ or $pp$
collisions. 

\subsection{$M_\perp$-dependence of YKP radii; transverse flow}
\label{sec5.2}

If the source expands rapidly and features large velocity gradients, the
``regions of homogeneity'' contributing to the correlation function
will be small. Their size will be inversely related to the velocity
gradients, scaled by a ``thermal smearing factor'' $\sqrt{T/M_\perp}$
which characterizes the width of the Boltzmann distribution.\cite{CSH95}
If one evaluates the expectation values (\ref{20a}-\ref{20c}) by
saddle point integration one finds for pairs with $Y=0$
 \begin{eqnarray}
 \label{Rsaddle}
    R_\perp^2 = R_*^2 \, , \qquad
    R_0^2 = (\Delta t_*)^2\, ,\qquad
    R_\parallel^2 = L_*^2 \, ,
 \end{eqnarray}
with
 \begin{eqnarray}
 \label{Rstar}
   {1\over R_*^2} &=& {1\over R^2} + {1\over R_{\rm flow}^2}\, , 
 \\
 \label{tstar}
   (\Delta t_*)^2 &=& (\Delta\tau)^2 + 
   2 \left( \sqrt{\tau_0^2 + L_*^2} - \tau_0 \right)^2 \, , 
 \\
 \label{Lstar}
   {1\over L_*^2} &=& {1\over (\tau_0\Delta\eta)^2} 
   + {1\over L_{\rm flow}^2}\, , 
 \end{eqnarray}
where $R_{\rm flow}$ and $L_{\rm flow}$ are the transverse and
longitudinal ``dynamical lengths of homogeneity'' due to the expansion
velocity gradients:
 \begin{eqnarray}
 \label{RH}
   R_{\rm flow}(M_\perp) &=& {R\over \eta_f}\, \sqrt{{T\over M_\perp}}
   = {1\over \partial \eta_t(r)/\partial r} \, \sqrt{{T\over M_\perp}}\, ,
 \\
 \label{LH}
   L_{\rm flow}(M_\perp) &=& \tau_0\, \sqrt{{T\over M_\perp}}
   = {1\over \partial{\cdot}u_l} \, \sqrt{{T\over M_\perp}}\, ,
 \end{eqnarray}
where $u_l$ is the longitudinal 4-velocity.

\begin{figure}[h]
\vspace*{10.5cm}
\includegraphics{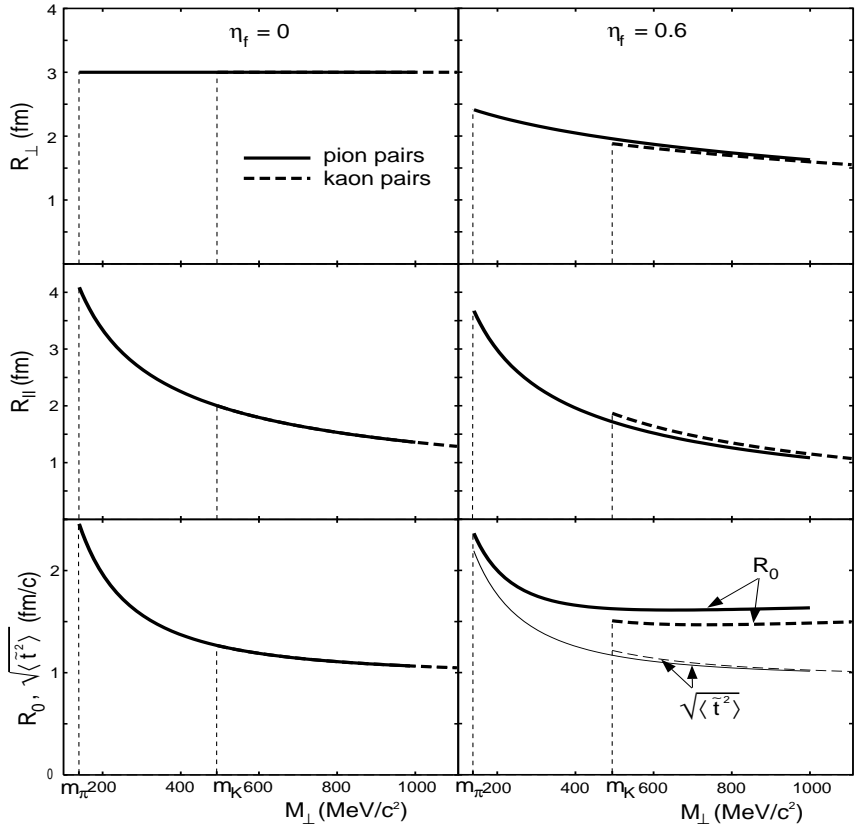}
\caption{The YKP radii $R_\perp$, $R_\parallel$, and $R_0$ (top
  to bottom) for zero transverse flow (left column) and for
  $\eta_f=0.6$ (right column), as functions of $M_\perp$ for pairs at
  $Y_{\rm cm}=0$. Solid (dashed) lines are for pions (kaons). The
  breaking of the $M_\perp$-scaling by transverse flow is obvious in
  the right column. For nonzero transverse flow $R_0$ also does not
  agree exactly with the effective source lifetime $\sqrt{\langle \tilde
  t^2\rangle}$ (lower right panel). Source parameters as in
  Fig.~\protect\ref{F1}. 
  \label{F2}}
\end{figure}

Thus, for expanding sources, the HBT radius parameters are generically 
decreasing functions of $M_\perp$. The slope of this decrease grows
with the expansion rate~\cite{WSH96,HTWW96} (this cannot be seen in
the saddle point approximated expressions above). Longitudinal
expansion affects mostly the longitudinal radius parameter
$R_\parallel$ and the temporal parameter $R_0$;\cite{HTWW96} the
latter is a secondary effect since particles from different points are
usually emitted at different times, and a decreasing longitudinal
homogeneity length thus also leads to a reduced effective duration of
particle emission (see lower panels in Fig.~\ref{F2}). 

The transverse radius parameter $R_\perp$ is invariant under
longitudinal boosts and thus not affected at all by longitudinal 
expansion (upper left panel in Fig.~\ref{F2}). It begins, however, to
drop as a function of $M_\perp$ if the source expands in the
transverse directions (upper right panel). Comparing the lower two 
left and right panels in Fig.~\ref{F2} one sees that the sensitivity
of $R_\parallel$ and $R_0$ to transverse flow is much
weaker.\cite{HTWW96} Transverse (longitudinal) flow thus mostly
affects the transverse (longitudinal) regions of homogeneity.

While longitudinal ``flow'' is not necessarily a signature for nuclear
collectivity but could be ``faked'' as discussed at the end of the
previous subsection, transverse flow is much more generic in this
respect: the ingoing channel has no transverse collective motion, and
the only mechanism imaginable for the creation of transverse
collective dynamics during the collision is multiple (re-)scattering
among the secondaries, leading ultimately to hydrodynamic transverse flow. 

Unfortunately, the observation of an $M_\perp$-dependence of $R_\perp$
by itself is not sufficient to prove the existence of radial transverse 
flow. It can also be created by other types of transverse gradients,
e.g. a transverse temperature gradient.\cite{CSH95,CL96,TH98a} To
exclude such a possibility one must check the $M_\perp$-scaling of the
YKP radii, i.e. the independence of the functions $R_i(M_\perp)$ 
($i=\perp,\parallel,0$) of the particle rest mass (which is not broken 
by temperature gradients). Since different particle species are 
affected differently by resonance decays, such a check further 
requires the elimination of resonance effects.  

\section{Resonance decays}
\label{sec6}

Resonance decays contribute additional pions at low $M_\perp$; these 
pions originate from a larger region than the direct ones, due to 
resonance propagation before decay. They cause an 
$M_\perp$-dependent modification of the HBT radii.  

Quantitative studies~\cite{Schlei,WH96} have shown that the resonances 
can be subdivided into three classes with different characteristic 
effects on the correlator: \\
(i) Short-lived resonances with lifetimes up to a few fm/$c$ do not 
propagate far outside the region of thermal emission and thus affect 
$R_\perp$ only marginally. They contribute to $R_0$ and $R_\parallel$
up to about 1 fm via their lifetime; $R_\parallel$ is larger if pion 
emission occurs later because for approximately boost-invariant expansion 
the longitudinal velocity gradient decreases as a function of time.\\ 
(ii) Long-lived resonances with lifetimes of more than several hundred fm/$c$ 
do not contribute to the measured correlation and thus only reduce the
correlation strength (the intercept at $q=0$), without changing the shape 
of the correlator.\footnote{A reduced correlation strength in the
  two-particle sector could also arise from partial phase coherence in
  the source.\cite{APW93} By comparing two- and three-particle
  correlations, the intercept-reducing effects of resonances can be
  eliminated, and the degree of coherence resp. chaoticity in the
  source can be unambiguously determined.\cite{HZ97}}
Decaying at large distances from their production point, they simulate
a very large source which contributes to the correlation signal only
for unmeasurably small relative momenta.\\  
(iii) Only the $\omega$ meson with its lifetime of 23.4 fm/$c$ does
not fall in either of these two classes and can thus distort the form
of the correlation function. It contributes a second bump at small $q$
to the correlator, giving it a non-Gaussian shape and thus
complicating~\cite{WH96} the extraction of HBT radii from a  
Gaussian fit. At small $M_\perp$ up to 10\% of the pions can come from
$\omega$ decays, and this fraction doubles effectively in the
correlator since the other pion can be a direct one; thus the effect
is not always negligible.   

\begin{figure}[h]
\vspace*{4.3cm}
\includegraphics{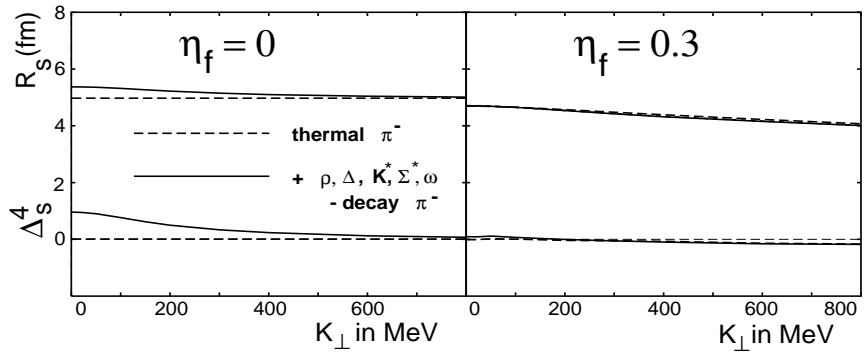}
\caption{The inverted $q$-variance $R_\perp$ and the kurtosis
  $\Delta_\perp$ (the index $s$ in the figure stands for ``sideward'')
  at $Y=0$ as functions of $K_\perp$. Left: $\eta_f=0$ (no transverse
  flow). Right: $\eta_f=0.3$. The difference between dashed and solid
  lines is entirely dominated by $\omega$ decays. Source parameters as
  in Fig.~\protect\ref{F1}, except for $R=5$ fm.
\label{F3}}
\end{figure}

In a detailed model study~\cite{WH96} we showed that resonance 
contributions can be identified through the non-Gaussian features in 
the correlator induced by the tails in the emission function resulting 
from resonance decays. To this end one computes the second and fourth 
order $q$-moments of the correlator.\cite{WH96} The second order 
moments define the HBT radii, while the kurtosis (the normalized
fourth order moments) provide a lowest order measure for the 
deviations from a Gaussian shape. We found~\cite{WH96} that, at least 
for the model (\ref{3.15}), a positive kurtosis can always be 
associated with resonance decay contributions (Fig.~\ref{F3}, left
panel). Strong flow also generates a non-zero, but small and
apparently always negative kurtosis (Fig.~\ref{F3}, right panel).
Any $M_\perp$-dependence of $R_\perp$ which is associated with a positive $M_\perp$-dependent kurtosis must therefore be regarded with suspicion; an 
$M_\perp$-dependence of $R_\perp$ with a vanishing or negative 
kurtosis, however, cannot be blamed on resonance decays.  

In our model, the first situation is realized for a source without 
transverse expansion (left panel of Fig.~1): At small $M_\perp$ 
the $\omega$ contribution increases $R_\perp$ by up to 0.5 fm while 
for $M_\perp > 600$ MeV it dies out. The effect on $R_\perp$ is small 
because the heavy $\omega$ moves slowly and doesn't travel very far 
before decaying. The resonance contribution is clearly visible in the 
positive kurtosis (lower curve). For non-zero transverse flow (right 
panel) there is no resonance contribution to $R_\perp$; this is 
because for finite flow the effective source size for the heavier 
$\omega$ is smaller than for the direct pions, and the $\omega$-decay 
pions thus always remain buried under the much more abundant direct 
ones. Correspondingly the kurtosis essentially vanishes; in fact, it 
is slightly negative, due to the weak non-Gaussian features induced by 
the transverse flow.  

\section{Opaque sources}
\label{sec7}

The emission function (\ref{3.15}) is only one of an infinity of
possible source pa\-ra\-me\-tri\-za\-tions. Its form permits, however,
easy implementation of most of the important features of the sources
created in heavy ion collisions. Still, there is one important
physical situation which cannot be parametrized by the formula
(\ref{3.15}): if the source emits particles not from the entire
volume, but only from a thin surface layer. This is how the sun
radiates photons, and this is also an often suggested picture for 
the slow hadronization of long-lived QGP blobs through a
deflagration-type strong first order transition.

The significance of such a phenomenon for HBT interferometry was
realized by Heiselberg and Vischer~\cite{HV96} who pointed out that
an effective emission region which is part of a thin surface layer has
a much smaller extension in the ``outward'' or $x$-direction than in
the ``sideward'' or $y$-direction. In other words, such ``opaque
sources'' have $\langle\tilde x^2 - \tilde y^2\rangle < 0$. Depending
on the degree of opacity (the thickness of the surface layer relative
to the source radius) this difference can be large and negative.
The authors pointed out~\cite{HV96} that this leads to the possibility
of a smaller ``outward'' than ``sideward'' HBT radius parameter in the
Pratt-Bertsch parametrization, even at $K_\perp = 0$. Recently
B. Tom\'a\v sik showed~\cite{TH98b} that in the YKP parametrization
opacity effects would show up even more spectacularly by a ``lifetime
parameter'' $R_0^2$ which would diverge to $-\infty$ in the limit
$K_\perp \to 0$ resp. $\beta_\perp \to 0$ (see Eq.~(\ref{20c})).

The source (\ref{3.15}) can be made opaque~\cite{HV96,TH98b} by
multiplying it by the factor $\exp\left( - \kappa l_{\rm eff} / 
\lambda) \right)$ where $\lambda$ is the mean free path and $\kappa
l_{\rm eff}$ is the effective travelling distance of the emitted
particle through matter in the source.\cite{TH98b}

\begin{figure}[h]
\vspace*{3.5cm}
\includegraphics{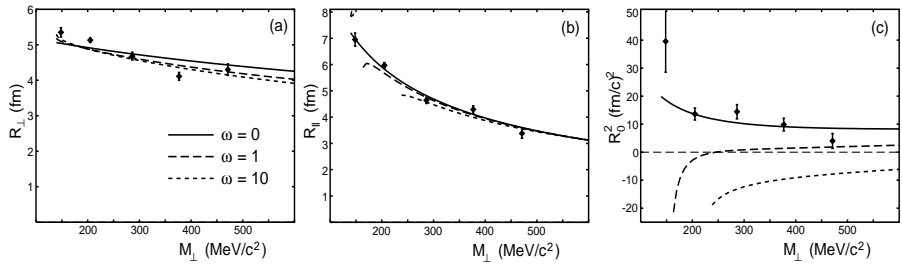}
\caption{The YKP radius parameters $R_\perp,R_\parallel$, and
  $R_0^2$ as functions of $M_\perp$ for pion pairs from 158 A GeV/$c$ 
  Pb+Pb collision~\protect\cite{NA49} at slightly forward pair rapidity
  $Y_{_{\rm CM}}=1.25$. Solid, dashed and dotted curves correspond to
  $R/\lambda = 0, 1,10$, respectively. (The corresponding pairs
  $(\eta_f,R)$ are (0.4,5.74\,fm), (0.345,4.83\,fm), and
  (0.215,3.35\,fm), respectively.\protect\cite{TH98b}) Other source
  parameters: $T$=120 MeV, $\tau_0$=8 fm/$c$, $\Delta\tau$=2 fm/$c$,
  $\Delta\eta$=1.3.
\label{F4}}
\end{figure}

Fig.~\ref{F4}c shows the ``temporal'' YKP radius parameter $R_0^2$ as
a function of $M_\perp$ for sources with different degrees of
opacity. For different opacities $R/\lambda$, the transverse source
parameters $R$ and $\eta_f$ were readjusted~\cite{TH98b} to give the
same measured~\cite{NA49} single particle slope. The crucial features
of opacity are clearly visible: the negative contribution $\sim
\langle\tilde x^2 - \tilde y^2\rangle$ in (\ref{20c}) drives $R_0^2$
to negative values at small $K_\perp$, and this happens the sooner the
shorter the mean free path $\lambda$, i.e. the thinner the surface
layer is. 

Fig.~\ref{F4}c implies that thin surfaces with $\lambda < R$ are
essentially excluded. Pion freeze-out occurs not in the form of surface
emission, but happens in bulk. 160 A GeV Pb+Pb collisions are thus
quite similar to the Early Universe: an early stage of {\em complete
  opaqueness} (here lasting for about 8 fm/$c$) is followed by a
rather {\em sudden} ($\Delta\tau \sim$1-2 fm/$c$) transition to
complete transparency. In both cases this transition is due to the
expansion and cooling of the system, causing a rapid increase of the
mean free path of the particles. In the Early Universe photon
decoupling is triggered by the recombination of electrons and ions
into neutral atoms; in heavy ion collisions at SPS energies pion
decoupling is caused mostly by the rapid cooling and dilution of the baryon density, since baryonic resonances with their strong coupling
to the pion channel provide the ``glue'' needed for keeping the system
close to local thermal equilibrium. 

\section{Analysis of Pb+Pb data}
\label{sec8}

Fig.~\ref{F4} shows a numerical fit~\cite{TH98b,WTH97} of the YKP radius
parameters, using the expressions (\ref{20a})-(\ref{20c}) with our
model source (\ref{3.15}), to data collected by the NA49 collaboration
in 158 A GeV/$c$ Pb+Pb collisions.\cite{NA49} The fit includes
resonance decay contributions to the single particle spectra, but not
to the 2-particle correlations. Based on sample calculations the
latter are, however, expected to be inside the systematic error of the
data. 

The width of the pion rapidity distribution is reproduced with $\Delta
\eta{=}1.3$. $\tau_0$ and $\Delta \eta$ are then fixed~\cite{TH98b} by
the magnitude of $R_\parallel$ and $R_0$. The magnitude of
$R_\perp(K_\perp=0)$ fixes the radius $R$ once $T$ and $\eta_f$ are
known. The latter are obtained from the $M_\perp$-dependence of
$R_\perp$, albeit not independently: essentially only the combination
$\eta_f \sqrt{M_\perp/T}$ (the velocity gradient divided by the
thermal smearing factor) can be extracted.\cite{CNH95,Sch96} This is
similar to the single particle spectra whose $M_\perp$-slopes
determine only an effective blushifted temperature,\cite{LH89} $T_{\rm
  eff} = T \sqrt{{1+ \bar v_f \over 1- \bar v_f}}$. The correlations
between $T$ and $\eta_f$ are, however, exactly opposite in the two
cases: for a fixed spectral slope $T$ must be decreased if $\eta_f$
increases while a fixed $M_\perp$-slope of $R_\perp$ requires
decreasing values of $\eta_f$ if $T$ is reduced.\cite{Sch96} {\em The
  combination of single-particle spectra and two-particle correlation
  thus allows for a separate determination of $T$ and $\eta_f$.} 

The fit in Fig.~\ref{F4} corresponds to an average transverse flow
velocity $\bar v_f{=}0.44$, combined with a freeze-out temperature of
about 120 MeV. Similar values were advocated by
K\"ampfer~\cite{kaempfer96} based on a simultaneous analysis of single
particle spectra of various different hadron species from Pb+Pb
collisions. 

Let us discuss in more detail the numbers resulting from this fit. 
First, the transverse size parameter $R\approx 6$ fm is surprisingly
large. At mid-rapidity even larger values ($R\simeq 7$ fm) are
required.\cite{NA49,Sch96} Resonance contributions are not expected to
reduce it by more than 0.5 fm.\cite{WH96} The transverse flow
correction to $R_\perp$ is appreciable, resulting in a visible
transverse homogeneity length of only about 5 fm at small $K_\perp$,
but even this number is large. $R{=}6$-7 fm corresponds to an
r.m.s. radius $r_{\rm rms} = \sqrt{\langle \tilde x^2 + \tilde y^2
  \rangle} \approx 8.5$-10 fm of the pion source, to be compared with an
r.m.s. radius $r_{\rm rms}^{\rm Pb} = 1.2 \times A^{1/3} * \sqrt{2/5}$
fm = 4.5 fm for the density distribution of the original Pb nucleus
projected on the transverse plane. This implies a transverse expansion
of the reaction zone by a linear factor $\simeq 2$. That we also find
a large transverse flow velocity renders the picture consistent. The
longitudinal size of the collision region at the point where the
pressure in the system began to drive the transverse expansion can be
estimated as follows: for the source to expand in, say, the
$y$-direction from $\sqrt{\langle y^2 \rangle}_{\rm initial}= 1.2\,
A^{1/3} /\sqrt{5}$ fm = 3.2 fm to $\sqrt{\langle y^2 \rangle}_{\rm
  final} = R \approx 6$\,fm with an average transverse flow velocity of at
most $\bar v = 0.44\, c$ (the freeze-out value determined from the fit
shown in Fig.~\ref{F4}) requires a time of at least $(6-3.2)/0.44$
fm/$c \simeq 6.5$ fm/$c$. Due to the selfsimilarity of the longitudinal
expansion the longitudinal dimension of the source grows linearly with
$\tau$. If the total expansion time until freeze-out is given by the
fit parameter $\tau_0 = 8$ fm/$c$, the source expanded in the $\simeq
6.5$ fm/$c$ during which there was transverse expansion by a factor
8/(8-6.5) = 8/1.5 $\simeq 5$ in the longitudinal direction. We
conclude that the fireball volume must have expanded by a factor $5 *
2^2 \approx 20$ between the onset of transverse expansion and
freeze-out! This is the clearest evidence for strong collective
dynamical behaviour in ultra-relativistic heavy-ion collisions so far.

The local comoving energy density at freeze-out can be estimated from 
the fitted values for $T$ and $\eta_f$. The thermal energy density of 
a hadron resonance gas at $T=120$ MeV and moderate baryon chemical 
potential is of the order of 100 MeV/fm$^3$. The large average 
transverse flow velocity of $\langle v_f \rangle \approx 0.44$ implies 
that about 25\% flow energy must be added in the lab frame. This 
results in an estimate of about $0.1$ GeV/fm$^3 \times 1.25 \times 20
\approx$ {\bf 2.5 GeV/fm}$^3$ for the energy density of the reaction
zone at the onset of transverse expansion.  This is well above the
critical energy density $\epsilon_{\rm cr} \leq 0.9$ GeV/fm$^3$
predicted by lattice QCD for deconfined quark-gluon matter.\cite{K96}
Whether this energy density was fully thermalized is, of course, a
different question. It must, however, have been accompanied by transverse
pressure (i.e. some degree of equilibration of momenta must have
occurred already before this point), because otherwise transverse
expansion could not have been initiated.   
                                        
\section{Conclusions}
\label{sec9}

I hope to have shown that

\begin{itemize}

\item
 two-particle correlation functions from heavy-ion collisions provide 
valuable information both on the geometry {\bf and} the dynamical 
state of the reaction zone at freeze-out;

\item
 a comprehensive and simultaneous analysis of single-particle spectra 
and two-particle correlations, with the help of models which provide
a realistic parametrization of the emission function, allows for an 
essentially complete reconstruction of the final state of the reaction 
zone, which can serve as a reliable basis for theoretical 
back-extrapolations towards the interesting hot and dense early stages 
of the collision;
 
\item
 simple and conservative estimates, based on the crucial new 
information from HBT measurements on the large transverse size of the 
source at freeze-out and using only energy conservation, lead to the 
conclusion that in Pb+Pb collisions at CERN, before the onset of 
transverse expansion, the energy density exceeded comfortably the 
critical value for the formation of a color deconfined state of quarks 
and gluons. There is, however, no evidence for long time delays due to
hadronization of the QGP, and pions freeze out in bulk rather
than from the surface of the collision fireball. This is in line with
lattice results which predict at most a {\em weakly} first order
confinement transition, and with other evidence~\cite{LTHSR95} for
rapid hadronization.

\end{itemize}

\noindent
 I would like to thank the organizers of CRIS98 for their hospitality
 and support. The results reported here were obtained in collaboration
 with D. Anchishkin, S. Chapman, P. Scotto, C. Slotta, B. Tom\'a\v sik,
 U. Wiedemann, Y.-F. Wu, and Q.H. Zhang. Our work was supported by
 grants from  DAAD, DFG, NSFC, BMBF, GSI and the Alexander von
 Humboldt Foundation.



\section*{References}

\begin{thebibliography}{99}

\bibitem{HBT}
  R. Hanbury Brown and R.Q. Twiss, {\em Phil. Mag.} {\bf 45}, 633
  (1954); and {\em Nature} {\bf 177}, 27 (1956); {\bf 178}, 1046 and
  1447 (1956).

\bibitem{GGLP}
  G. Goldhaber, S. Goldhaber, W. Lee, and A. Pais, \PR {\bf 120}, 300
  (1960).

\bibitem{3pi}
  J. Schmidt-S\"orensen, this volume, and references therein.

\bibitem{HZ97}
  U. Heinz and Q.H. Zhang, \PRC {\bf 56}, 426 (1997); H. Heiselberg
  and A. Vischer, nucl-th/9707036.

\bibitem{LH89}
  K.S. Lee and U. Heinz, \ZPC {\bf 43}, 425 (1989);
  K.S. Lee, U. Heinz and E. Schnedermann, \ZPC {\bf 48}, 525 (1990);
  E. Schnedermann and U. Heinz, \PRL {\bf 69}, 2908 (1992);
  E. Schnedermann, J. Sollfrank and U. Heinz, \PRC {\bf 48}, 2462 (1993);
  E. Schnedermann and U. Heinz, \PRC {\bf 50}, 1675 (1994);
  U. Heinz, in {\it Hot Hadronic Matter: Theory and Experiment}, 
  (J. Letessier {\it et al.}, eds.), NATO ASI Series {\bf B346}
  (Plenum, New York, 1995), p.~413.

\bibitem{Stach94}
  J. Stachel {\it et al.}, [E814 Collaboration], \NPA {\bf 566}, 183c
  (1994);   
  P. Braun-Munzinger {\it et al.}, \PLB {\bf 344}, 43 (1995).

\bibitem{QM96}
  See proceedings of {\it Quark Matter 96}, \NPA {\bf 610} (1996),
  and of  {\it Quark Matter 97}, \NPA (1998), in press.

\bibitem{HB95}
  M. Herrmann and G.F. Bertsch, \PRC {\bf 51}, 328 (1995).

\bibitem{CSH95} 
  S. Chapman, P. Scotto and U. Heinz, \PRL {\bf 74}, 4400 (1995);
  and {\em Heavy Ion Physics} {\bf 1}, 1 (1995).

\bibitem{CNH95}
  S. Chapman, J.R. Nix and U. Heinz, \PRC {\bf 52}, 2694 (1995).

\bibitem{P84}
  S. Pratt, \PRL {\bf 53}, 1219 (1984); \PRD {\bf 33}, 1314 (1986).

\bibitem{MS88}
  A.N. Makhlin and Yu.M. Sinyukov, \ZPC {\bf 39}, 69 (1988).

\bibitem{AS95}
  S.V. Akkelin and Y.M. Sinyukov, \PLB {\bf 356}, 525 (1995).

\bibitem{CL96}
  T. Cs\"org\H o and B. L\"orstad, \PRC {\bf 54}, 1396 (1996).
 
\bibitem{TH98a}
  B. Tom\'a\v sik and U. Heinz, \EPJC (23. Feb. 1998) (online
  publication) [nucl-th/9707001].

\bibitem{WSH96}
  U.A. Wiedemann, P. Scotto and U. Heinz, \PRC {\bf 53}, 918 (1996).

\bibitem{HTWW96} 
  U. Heinz {\it et al.}, \PLB {\bf 382}, 181 (1996); 
  Wu Y.-F. {\it et al.}, \EPJC {\bf 1}, 599 (1998).

\bibitem{Schlei}
  B.R. Schlei {\it et al.}, \PLB {\bf 293}, 275 (1992); J. Bolz {\it et 
  al.}, {\it ibid.} {\bf 300}, 404 (1993); \PRD {\bf 47}, 3860 (1993).

\bibitem{WH96}
  U.A. Wiedemann and U. Heinz, \PRC {\bf 56}, R610 (1997);
  and \PRC {\bf 56}, 3265 (1997).

\bibitem{He96}
  U. Heinz, in {\it Correlations and Clustering Phenomena in Subatomic 
  Physics}, (M.N. Harakeh {\it et al.}, eds.), NATO ASI Series {\bf
  B359} (Plenum, New York, 1997), p.~137.

\bibitem{GKW79}
  M. Gyulassy, S.K. Kauffmann, and L.W. Wilson, \PRC {\bf 20}, 2267
  (1979).

\bibitem{S73} 
  E. Shuryak, \PLB {\bf 44}, 387 (1973); {\em Sov. J. Nucl. Phys.} 
  {\bf 18}, 667 (1974).

\bibitem{CH94}
  S. Chapman and U. Heinz, \PLB {\bf 340}, 250 (1994).

\bibitem{AHR97}
  D. Anchishkin, U. Heinz, and P. Renk, \PRC {\bf 57}, 1428 (1998).

\bibitem{zhang97}
  Q.H. Zhang, U.A. Wiedemann, C. Slotta, and U. Heinz, \PLB {\bf 407},
  33 (1997).

\bibitem{Wetal}
  U.A. Wiedemann {\it et al.}, \PRC {\bf 56}, R614 (1997).

\bibitem{ZC98}
  T. Cs\"org\H o and J. Zim\'anyi, \PRL {\bf 80}, 916 (1998); 
  J. Zim\'anyi and T. Cs\"org\H o, hep-ph/9705432.

\bibitem{W98}
  U.A. Wiedemann, nucl-th/9801009

\bibitem{MP98}
  D. Pelte and H. Merlitz, this volume, and references therein
  [nucl-th/9806049].

\bibitem{Barz}
  H.W. Barz, \PRC {\bf 53}, 2536 (1996); S. Pratt and M.B. Tsang, \PRC
  {\bf 36}, 2390 (1987).

\bibitem{AH98}
  D. Anchishkin and U. Heinz, in preparation.

\bibitem{HW98}
  U. Heinz and U.A. Wiedemann, in preparation.

\bibitem{brown97}
  D.A. Brown and P. Danielewicz, \PLB {\bf 398}, 252 (1997); and \PRC
  {\bf 57}, 2474 (1998).

\bibitem{He97}
  U. Heinz, in: Proceedings of the 5th International Workshop on {\it
    Relativistic Aspects of Nuclear Physics}, (T. Kodama {\it et al.},
  eds.), World Scientific, Singapore, 1998, in press [nucl-th/9710065].

\bibitem{EGHW98}
  J. Ellis, K. Geiger, U. Heinz and U.A. Wiedemann, in preparation. 

\bibitem{APW93}
  I.V. Andreev, M. Pl\"umer, and R.M. Weiner, {\em Int. J. Mod. Phys.}
  A {\bf 8}, 4577 (1993).

\bibitem{HV96}
  H. Heiselberg and A.P. Vischer, \EPJC {\bf 2}, 593 (1998).

\bibitem{TH98b}
  B. Tom\'a\v sik and U. Heinz, nucl-th/9805016, subm. to \EPJC.

\bibitem{NA49}
  H. Appelsh\"auser {\it et al.} (NA49 Collaboration), \EPJC {\bf 2},
  661 (1998); P.G. Jones {\it et al.} (NA49 Collaboration), 
  \NPA {\bf 610}, 188c (1996); R. Ganz, this volume.

\bibitem{alber95}
  T. Alber {\em et al.} (NA35 Collaboration), \ZPC {\bf 66}, 77 (1995).

\bibitem{WTH97}
  U.A. Wiedemann, B. Tom\'a\v sik, and U. Heinz, in {\it Quark Matter
    97}, \NPA (1998), in press [nucl-th/9801017].

\bibitem{Sch96}
  S. Sch\"onfelder, PhD thesis, MPI f\"ur Physik, M\"unchen (1996).  

\bibitem{kaempfer96}
  B. K\"ampfer, nucl-th/9612336, and private communication.

\bibitem{K96}
  E. Laermann, \NPA {\bf 610}, 1c (1996).

\bibitem{LTHSR95}
  J. Letessier, A. Tounsi, U. Heinz, J. Sollfrank, and J. Rafelski, 
  \PRC {\bf 51}, 3408 (1995).

\end{thebibliography}
\end{document}